\documentclass[12pt]{article} %[reprint,aps]{revtex4-1}
\usepackage{epsfig, amsmath, amssymb}
\usepackage{graphicx,psfrag} 
\newcommand{\be}{\begin{equation}}
\newcommand{\ee}{\end{equation}}
\newcommand{\ben}{\begin{equation}}
\newcommand{\een}{\end{equation}}
\newcommand{\bea}{\begin{eqnarray}}
\newcommand{\eea}{\end{eqnarray}}
\newcommand{\bA}{\begin{array}}
\newcommand{\eA}{\end{array}}
\newcommand{\bc}{\begin{center}}
\newcommand{\ec}{\end{center}}

\newcommand{\ra}{\rightarrow}
\newcommand{\del}{\partial}

\newcommand{\ie}{{\it i.e.}}
\newcommand{\eg}{{\it e.g.}}

\newcommand{\vx}{{\vec x}}

\newcommand{\cO}{{\cal O}}
\newcommand{\cI}{{\cal I}}

\begin{document}

%\ifeprint
%\fi

\begin{titlepage}
%\vspace{30mm}

\bc

%%\hfill  {TIFR/TH/09-12} \\
\hfill % {\tt arXiv:0909.4731 [hep-th]} 
\\         [40mm]
%%X\vfill

{\Huge de Sitter space and 
 \\ [2mm] extremal surfaces for spheres} 
\vspace{16mm}

{\large K.~Narayan} \\
\vspace{3mm}
{\small \it Chennai Mathematical Institute, \\}
{\small \it SIPCOT IT Park, Siruseri 603103, India.\\}
%%{\small Email: \ narayan@cmi.ac.in}\\

\ec
\medskip
\vspace{40mm}

\begin{abstract}
Following arXiv:1501.03019 [hep-th], we study de Sitter space and
spherical subregions on a constant boundary Euclidean time slice of
the future boundary in the Poincare slicing.  We show that as in that
case, complex extremal surfaces exist here as well: for even boundary
dimensions, we isolate the universal coefficient of the
logarithmically divergent term in the area of these surfaces. There
are parallels with analytic continuation of the Ryu-Takayanagi
expressions for holographic entanglement entropy in $AdS/CFT$. We then
study the free energy of the dual Euclidean CFT on a sphere
holographically using the $dS/CFT$ dictionary with a dual de Sitter
space in global coordinates, and a classical approximation for the
wavefunction of the universe.  For even dimensions, we again isolate
the coefficient of the logarithmically divergent term which is
expected to be related to the conformal anomaly. We find agreement
including numerical factors between these coefficients.
\end{abstract}

\end{titlepage}

%\newpage 
%{\tiny %footnotesize
%\begin{tableofcontents}
%\end{tableofcontents}
%}

%\vspace{5mm}

\section{Introduction}

Generalizations of gauge/gravity duality \cite{Maldacena:1997re,
Gubser:1998bc,Witten:1998qj,Aharony:1999ti} to de Sitter space, or
$dS/CFT$ \cite{Strominger:2001pn,Witten:2001kn,Maldacena:2002vr},
involve a hypothetical dual Euclidean CFT on the future timelike infinity 
${\cal I}^+$ boundary. The late time wavefunction of the universe with 
appropriate boundary conditions is equated with the partition function 
of the dual CFT. Further work on $dS/CFT$ including higher spin 
realizations appears in \eg\ 
\cite{Harlow:2011ke,Anninos:2011ui,Ng:2012xp,Das:2012dt,Anninos:2012ft,
Das:2013qea,Banerjee:2013mca,Das:2013mfa}.

Ideas pertaining to entanglement entropy have been of great interest
in recent times. In $AdS/CFT$, the Ryu-Takayanagi prescription
\cite{Ryu:2006bv,Ryu:2006ef} (see \cite{HEEreview,HEEreview2} for
reviews) maps entanglement entropy of a field theory subsystem to the
area (in Planck units) of a bulk minimal surface (more generally
extremal surface \cite{HRT}) anchored at the subsystem interface and
dipping into the bulk, in the gravity approximation.  Similar ideas
were explored in \cite{Narayan:2015vda} in de Sitter space with a view
to exploring entanglement entropy in the dual CFT with $dS/CFT$ in
mind. For strip-shaped subregions on a constant Euclidean time slice
of the future boundary, it was found that the area of certain complex
extremal surfaces has structural resemblance with entanglement 
entropy of dual Euclidean CFTs (reviewed in sec.~2). The coefficients 
of the leading divergent ``area law'' terms in $dS_{d+1}$ resemble 
the central charges ${\cal C}_d\sim i^{1-d} {R_{dS}^{d-1}\over G_{d+1}}$ 
of the CFT$_d$s appearing in the $\langle T T\rangle$ correlators in
\cite{Maldacena:2002vr}. The areas of these surfaces obtained thus
essentially amount to analytic continuation from the Ryu-Takayanagi
expressions for holographic entanglement entropy in $AdS/CFT$. Note
that the areas are in general not real-valued or positive definite and
are distinct from the entanglement entropy of bulk fields in de Sitter
space \eg\ \cite{Maldacena:2012xp}.

Towards exploring this further, we study spherical subregions in this
paper. As in \cite{Narayan:2015vda}, similar complex extremal surfaces
can be shown to exist in this case too (sec.~3). The area of these
surfaces exhibits a similar leading area law divergence as well as
subleading terms: for $dS_{d+1}$ with even $d$, this includes a
logarithmically divergent term whose coefficient is analogous to the
``universal'' terms in $AdS/CFT$ related to the conformal anomaly
\cite{Ryu:2006ef}.  In the present case also, we expect this anomaly
to arise in the free energy of the CFT on a curved space. With this in
mind, we then calculate in the present case (sec.~4), the free energy
of the Euclidean CFT on a sphere holographically using the $dS/CFT$
dictionary $Z_{CFT} = \Psi$\ \cite{Maldacena:2002vr} with an auxiliary
de Sitter space in global coordinates (whose constant time slices are
spheres). In the classical regime, we approximate the wavefunction of
the universe $\Psi$ in terms of the bulk action $S$ of this auxiliary
de Sitter space: this gives $-F=\log Z_{CFT} = \log \Psi\sim iS$.  We
find precise agreement between the coefficients of the logarithmic
terms in the complex extremal surfaces and those in the free energy
via the wavefunction of the universe, including numerical
factors. Since the coefficient of the logarithmic term in the free
energy is related to the trace anomaly for any CFT, this supports the
idea that the area of these complex extremal surfaces encodes
entanglement entropy of the dual Euclidean CFT in $dS/CFT$, as we
discuss in sec.~5.

\section{Reviewing de Sitter extremal surfaces:\ strips}

Here we review the study \cite{Narayan:2015vda} of bulk de Sitter
extremal surfaces anchored over strip-shaped subregions on the future 
boundary ${\cal I}^+$ in the Poincare slicing.
de Sitter space $dS_{d+1}$ in the Poincare slicing or planar coordinate 
foliation is given by the metric
\be\label{dSpoinc}
ds^2 = {R_{dS}^2\over\tau^2} (-d\tau^2 + dw^2 + d\sigma_{d-1}^2)\ ,
\ee
where half of the spacetime, \eg\ the upper patch, has ${\cal I}^+$ 
at $\tau = 0$ and a coordinate horizon at $\tau=-\infty$.  This may be 
obtained by analytic continuation of a Poincare slicing of $AdS$, 
\be\label{AdStodS}
z \rightarrow -i\tau\ ,\qquad R_{AdS} \rightarrow -iR_{dS}\ ,\qquad 
t\ra -iw\ ,
\ee
where $w$ is akin to boundary Euclidean time, continued from time in 
$AdS$ (with $z$ the bulk coordinate).
The dual Euclidean CFT is taken as living on the future $\tau=0$
boundary ${\cal I}^+$.  We assume translation invariance with respect
to the boundary Euclidean time direction $w$, and consider a
subregion on a $w=const$ slice of ${\cal I}^+$. One might imagine that
tracing out the complement of this subregion then gives entropy in
some sense stemming from the information lost.  In the bulk, we study
de Sitter extremal surfaces on the $w=const$ slice, analogous to the
Ryu-Takayanagi prescription in $AdS/CFT$. Operationally these extremal
surfaces begin at the interface of the subsystem (or subregion) and
dip into the bulk time direction.

For a strip-shaped subregion on ${\cal I}^+$\ (with width say along 
$x$), parametrizing the spatial part in (\ref{dSpoinc}) as 
$d\sigma_{d-1}^2=\sum_{i=1}^{d-1} dx_i^2$, the $dS_{d+1}$ area functional 
on a $w=const$ slice is
\be\label{SdS-EEdS}
S_{dS} = {R_{dS}^{d-1} V_{d-2}\over 4G_{d+1}} 
\int {d\tau\over\tau^{d-1}} \sqrt{\Big({dx\over d\tau}\Big)^2-1\ }\ ,
\qquad\quad {\dot x}^2 = {-A^2\tau^{2d-2}\over 1-A^2\tau^{2d-2}}\ ,
\ee
with ${dx\over d\tau} \equiv {\dot x}$, and the constant $A^2$ is the 
conserved quantity obtained in the extremization. First let us 
consider $dS_4$ with the bulk time parametrized by real $\tau$ with 
the range $-\infty<\tau<0$ and correspondingly real surfaces, described 
in \cite{Narayan:2015vda}. These surfaces are obtained by taking 
$A^2<0$, which gives ${\dot x}^2 = {|A|^2\tau^{4}\over 1+|A|^2\tau^{4}}$~.
%\footnote{
For real $\tau$, we note that a crucial sign difference from the 
$AdS$ case implies the absence of a turning point where 
${\dot x}\ra\infty$. These are timelike surfaces with ${\dot x}^2\ra 0$ 
as $\tau\ra 0$ (anchored on the subregion boundary at ${\cal I}^+$, 
dipping into the past): as $|\tau|\ra\infty$, we have ${\dot x}^2\ra 1$ 
asymptoting to a null surface. An extremal surface can then be 
constructed by taking two half-extremal-surfaces bending ``inward'' 
and joining them with a cusp (see Fig.~1 in \cite{Narayan:2015vda}). 
Since these are real surfaces, it is natural to take 
$S_{dS} = {R_{dS}^{d-1} V_{d-2}\over 4G_{d+1}} \int {d\tau\over\tau^{d-1}} 
\sqrt{1-{\dot x}^2\ }$ as for a real timelike surface. 
The area decreases as $|A|^2$ increases: as $|A|^2\ra\infty$, these 
real surfaces become null with ${\dot x}^2\ra 1$ and are the analogs 
of surfaces with minimal (zero) area. They are restrictions (to a 
boundary Euclidean time slice) of the boundary of the past lightcone 
wedge of the boundary subregion, with vanishing area, and no bearing
on entanglement. One can also consider half-extremal-surfaces bending
``outward'' from the interface: again minimal area surfaces are null
with zero area.  Taking $A^2=0$ gives disconnected surfaces
$x(\tau)=const$, again with no turning point: it is then natural to
take $\tau$ to extend all the way to $|\tau|\ra\infty$ which gives
$S_{dS} \sim {R_{dS}^{d-1}\over G_{d+1}} {V_{d-2}\over\epsilon^{d-2}}$ with 
no cutoff-independent terms (encoding the interesting finite 
size-dependent part of entanglement). Thus real $dS$ extremal surfaces 
do not give interesting entanglement structure. Real codim-1 surfaces 
have similar behaviour.

With $dS/CFT$ in mind, we now consider $A^2>0$: this gives a complex 
surface\footnote{Complex geodesics and surfaces have also appeared 
in \eg\ \cite{Fidkowski:2003nf,Fischetti:2014uxa}.}. For $dS_4$, we 
have $x(\tau)\sim \pm iA\tau^3 + x(0)$ as $\tau\sim 0$, so that 
$x(\tau)$ representing a spatial direction in the CFT is real-valued 
only if $\tau$ is pure imaginary. More generally, requiring that the 
width $\Delta x=l$ be real-valued  suggests that $\tau$ takes imaginary 
values, parametrized as $\tau=iT$ with $T$ real. There is now a 
turning point $\tau_*={i\over \sqrt{A}}$ which is the ``deepest'' 
location this (smooth) complex surface dips upto in the bulk (with 
$|{\dot x}|\ra\infty$): in other words, $|\tau|\leq |\tau_*|$. The 
width condition $\Delta x = l$ can be shown to give $\tau_*=il$, so 
that large width $l\ra\infty$ implies $|\tau_*|\ra\infty$.
It is worth noting that the $\tau$-parametrization here lies outside 
the original de Sitter parametrization where $\tau$ was a real 
coordinate: instead for this complex solution to the extremization 
(with width $l$), $\tau$ runs along the imaginary axis, ranging 
from $i\epsilon$ to $il$. The corresponding surface 
$x(\tau)$ does not directly correspond to any real bulk $dS_4$ 
subregion: instead, complexified $\tau$ suggests an effective 
analytic continuation of (\ref{dSpoinc}) to Euclidean $AdS$ and a 
corresponding extremal surface.\ For even $d$, similar 
analysis can be done \cite{Narayan:2015vda} with complex surface 
saddle points of the area functional arising by taking $A^2<0$ and 
similar paths $\tau=iT$ (the details are different from $dS_4$). 
The area of these surfaces is
\be\label{EEdSd+1}
S_{dS} = -i {R_{dS}^{d-1}\over 4G_{d+1}} V_{d-2} 
\int_{\tau_{UV}}^{\tau_*} {d\tau\over\tau^{d-1}} 
{1\over\sqrt{1-(-1)^{d-1}A^2\tau^{2d-2}}}\ =\
i^{1-d} {R_{dS}^{d-1}\over 2G_{d+1}} V_{d-2} 
 \Big({1\over\epsilon^{d-2}} - c_d {1\over l^{d-2}}\Big) ,\  
\ee
where $\tau_{UV}=i\epsilon$ and $\tau_*=il$, and the integral is 
as in $AdS$ (with corresponding constant $c_d$). Note that here we 
have used the relation $\tau_{UV}=i\epsilon$ for the ultraviolet 
cutoff in the dual Euclidean field theory suggested by previous 
investigations in $dS/CFT$ (see \eg\ \cite{Maldacena:2002vr,
Harlow:2011ke,Anninos:2011ui,Das:2013qea}) with time evolution 
mapping to renormalization group flow.

$S_{dS}$ in (\ref{EEdSd+1}) bears structural resemblance to 
entanglement entropy in a dual CFT with central charge 
${\cal C}_d\sim i^{1-d} {R_{dS}^{d-1}\over G_{d+1}}$.
The first term $S_{dS}^{div}\sim i^{1-d} {R_{dS}^{d-1}\over G_{d+1}} 
{V_{d-2}\over\epsilon^{d-2}}$ resembles an area law divergence 
\cite{AreaLaw1,AreaLaw2}, proportional to the area of the interface
between the subregion and the environment, in units of the ultraviolet
cutoff. It appears independent of the shape of the subregion,
expanding (\ref{SdS-EEdS}) and assuming that ${\dot x}$ is small 
near the boundary $\tau_{UV}$. Written as 
${\cal C}_d {V_{d-2}\over\epsilon^{d-2}}$, we see that it is also 
proportional to the central charge 
${\cal C}_d\sim i^{1-d} {R_{dS}^{d-1}\over G_{d+1}}$ representing 
the number of degrees of freedom in the dual (non-unitary) CFT: 
these arose in the $\langle T T\rangle$ correlators obtained in 
\cite{Maldacena:2002vr}. In $dS_4$, the central charge 
${\cal C}\sim -{R_{dS}^2\over G_4}$ is real and negative, while 
in $dS_3, dS_5$, it is imaginary. The second term is a finite 
cutoff-independent piece.
Unlike $dS_4$, note that $S_{dS}$ in $dS_{d+1}$ with even $d$ is not 
real-valued: \eg\ in $dS_3$, we obtain
$S_{dS}\sim  -i{R_{dS}\over G_3} \log {l\over\epsilon}$ while in 
$dS_5$, we have $S_{dS} \sim\ i {R_{dS}^{3}\over G_{5}} V_{2} 
({1\over\epsilon^2}-c_4 {1\over l^2})$. Similar complex surfaces 
can be studied in the $dS$ black brane \cite{Das:2013mfa} which are 
dual to the CFT at uniform energy density: then the finite part 
resembles an extensive thermal entropy, again with a coefficient 
central charge as above. It is interesting to note 
that a replica calculation of entanglement entropy in a free 3d 
$Sp(N)$ theory for the half-plane \cite{Sato:2015tta} gives 
behaviour similar to the leading area law divergence here (although 
the $Sp(N)$ theory is dual to the higher spin $dS_4$ theory 
\cite{Anninos:2011ui} and it is unclear if geometric objects such 
as extremal surfaces are of relevance). 

While there is structural resemblance with entanglement entropy, there
are questions. Since these are bulk complex extremal surfaces,
changing the sign in the square root branch in (\ref{SdS-EEdS})
introduces an overall $\pm i$ factor. Fixing this in (\ref{EEdSd+1})
as $-i$ makes the leading divergence to be of the form of the area law
${\cal C}_d {V_{d-2}\over\epsilon^{d-2}}$ with ${\cal C}_d$ the
central charges in \cite{Maldacena:2002vr}. The resulting expressions
are corroborated by and essentially amount to analytic continuation
from the Ryu-Takayanagi expressions for holographic entanglement
entropy in $AdS/CFT$.  While this is suggestive, it would be useful to
explore this further with a view to associating these complex extremal
surfaces and corresponding area with entanglement entropy in $dS/CFT$.

Here we study spherical subregions and the corresponding complex
extremal surfaces. For $dS_{d+1}$ with even $d$, there is a term in
the area with logarithmic dependence on the cutoff $\epsilon$ whose
coefficient can be compared with that obtained from the conformal
anomaly appearing in the free energy of the CFT on a sphere
holographically using $dS/CFT$. We obtain agreement between both
sides: this vindicates the signs we have used above in defining these
complex surfaces, and analytic continuation.

\section{Spherical extremal surfaces in de Sitter space}

Building on \cite{Narayan:2015vda} for strip-shaped subregions, here 
we consider spherical subregions on the boundary ${\cal I}^+$, with
radius $l$ parametrized as $0\leq r\leq l$. 
Since we are interested in spherical entangling surfaces, we will 
parametrize $d\sigma_{d-1}^2$ in (\ref{dSpoinc}) in polar coordinates. 
Then the $w=const$ surface (\ie\ a constant boundary Euclidean time 
surface) is a bulk $d$-dim subspace with metric
\be\label{dSPoinc-wslice}
ds^2 = {R_{dS}^2\over\tau^2} \Big(-d\tau^2 + dr^2 + r^2 d\Omega_{d-2}^2\Big) .
\ee
The bulk surface on the $w=const$ slice bounding this subregion and 
dipping into the $\tau$-direction is bulk codim-2: let us parametrize 
this as $r=r(\tau)$. Its area functional in Planck units is
\be\label{RTdS01}
S_{dS} = {1\over 4G_{d+1}} 
\int \prod_{i=1}^{d-2} {R_{dS} r d\Omega_i\over\tau} {R_{dS}\over\tau} 
\sqrt{dr^2-d\tau^2} = {R_{dS}^{d-1} \Omega_{d-2}\over 4G_{d+1}} 
\int {d\tau\over\tau^{d-1}} r^{d-2} \sqrt{\Big({dr\over d\tau}\Big)^2-1~}\ .
\ee
The variational equation of motion for an extremum\ 
${\del\over\del\tau} ({\del {\cal L}\over\del {\dot r}}) = 
{\del {\cal L}\over \del r}$\ gives
\be\label{EOM}
{\del\over\del\tau} \Big( {r^{d-2}\over\tau^{d-1}} 
{{\dot r}\over\sqrt{{\dot r}^2-1}} \Big) = {d-2\over \tau^{d-1}} 
r^{d-3} \sqrt{{\dot r}^2-1} \ ,
\ee
where ${dr\over d\tau}\equiv {\dot r}$.\ It can be seen that 
\be\label{sphExtSur-1}
r(\tau) = \sqrt{l^2 + \tau^2}
\ee
is an extremal surface that solves (\ref{EOM}), thus extremizing 
$S_{dS}$, using
\be\label{sphExtSur-2}
{\dot r} = {\tau\over \sqrt{l^2+\tau^2}}\ ,\qquad 
{\dot r}^2 - 1 = {-l^2\over l^2+\tau^2}\ .
\ee
This satisfies the boundary conditions which require the surface to 
be anchored at the subregion interface, \ie\ $r\ra l$ as $\tau\ra 0$.
Unlike the strip case, there are no parameters for the surface in 
this case\footnote{For $d=2$, with just one spatial dimension, there 
is no difference between a strip and a sphere so the analysis is 
similar to that in \cite{Narayan:2015vda}. In detail, from (\ref{EOM}) 
we have    %${{\dot r}\over\sqrt{{\dot r}^2 -1}} = B\tau$ or 
${\dot r}^2 = {-B^2\tau^2\over 1-B^2\tau^2}$. With $B^2<0$, these are 
real surfaces parametrized as 
$r(\tau)=\pm\sqrt{\tau^2+(1/B^2)} + C$ subject to the boundary 
conditions \eg\ $r\ra \pm {l\over 2}$ as $\tau\ra 0$. ``Inward'' 
bending half-extremal-surfaces appropriately joined (Fig.~1 in 
\cite{Narayan:2015vda}) can be constructed asymptoting to null 
surfaces with zero area. Alternatively ``outward'' bending surfaces 
which extend all the way to $|\tau|\ra\infty$ are represented by 
\eg\ the two half-surfaces $r_L=-\sqrt{(l^2/4)+\tau^2} ,\ 
r_R=\sqrt{(l^2/4)+\tau^2}$, with area $S_{dS}= 2 {R_{dS}\over 4G_3} 
\int_{-\infty}^{-\epsilon} {d\tau\over\tau} {l/2\over\sqrt{(l^2/4)+\tau^2}} 
= {R_{dS}\over 2G_3} \log {l\over 2\epsilon}$~.} 
for $d>2$.
We see from (\ref{sphExtSur-1}), (\ref{sphExtSur-2}), that for $\tau$ 
real, there is no bulk turning point where ${dr\over d\tau}\ra\infty$ 
with the surface turning around smoothly: instead the surface 
asymptotically approaches $r^2\ra\tau^2$. Furthermore this surface 
has $r(\tau)\geq l$ whereas all interior points within the subregion 
satisfy $0\leq r\leq l$ with $r\ra l$ near $\tau\sim 0$, so that this 
surface bends ``outwards'' from the subregion boundary. This is a real 
timelike surface with ${\dot r} \leq 1$.   %\footnote{\label{Ft1} 
It is then more natural to consider, rather than (\ref{RTdS01}), 
the area as\ $S_{dS} = {R_{dS}^{d-1} \Omega_{d-2}\over 4G_{d+1}} 
\int {d\tau\over\tau^{d-1}} r^{d-2} \sqrt{1-{\dot r}^2}$ which is 
real-valued. Since the surface does not ``end'' at any finite 
$\tau$, we consider the whole $\tau$-range and obtain\ 
$S_{dS}={R_{dS}^{d-1} \Omega_{d-2} l \over 4G_{d+1}} 
\int_{-\infty}^{-\epsilon} {d\tau\over\tau^{d-1}} (l^2+\tau^2)^{(d-3)/2}$ ,
taking $\tau_{UV}=-\epsilon$. This gives 
%{\pi R_{dS}^2\over 2G_4} {l\over\epsilon} 
$S_{dS}= {R_{dS}^2\over 4G_4} {A_1\over\epsilon}$\ [for $dS_4$, 
with $A_1=2\pi l$ the interface area of the circular subregion],\ and
$S_{dS} = {R_{dS}^3\over 8G_5} {A_2\over\epsilon^2} + 
{\pi R_{dS}^3\over 2G_5} \log {l\over\epsilon}$ [for $dS_5$, with 
$A_2=4\pi l^2$ the 2-sphere interface area]. Note that there are 
no interesting finite cutoff-independent pieces for these surfaces 
since those contributions die at $|\tau|\ra\infty$.

From the point of view of the dual Euclidean CFT, we expect that the
central charge coefficients in the extremal surface area (interpreted
as entanglement entropy) must match those of the CFT. For the leading
area law divergence (which is not sensitive to the detailed geometry
of the subregion) of a spherical subregion, the scaling must be the
same as for the strip, matching the central charges obtained in
\cite{Maldacena:2002vr} which are negative or imaginary. In addition
there are expected to be universal coefficients for the sphere case
which should match CFT anomaly coefficients.  Further we would also
intuitively expect that there exist interesting finite
cutoff-independent parts which are size-dependent measures of
entanglement entropy from the CFT point of view. Given these
expectations, the real surfaces discussed above are unsatisfactory.

This suggests that we consider imaginary $\tau$ parametrized as 
$\tau=iT$ with $T$ real, as for the strip case. Thus (\ref{sphExtSur-1})
becomes
\be\label{sphExtSur-3}
r^2 = l^2 - T^2\quad \Rightarrow\quad r_{min}=0\ \ {\rm at\ the\ 
turning\ point}\ \
\tau_*=il \qquad \Rightarrow\quad \Delta r = l\ .
\ee
Now $r(\tau)$ maps each point on the surface directly to a
corresponding real-valued spatial location within the subregion in the
dual CFT (\ie\ the surface bends ``inward''). We require that the
boundary subregion radial parameter $r$ be real-valued in
(\ref{sphExtSur-1}) since this represents a spatial direction in the
CFT: this excludes more general paths in complex $\tau$-space. The
range of $\tau$ is now restricted, and the subregion size given by
$\Delta r \equiv r_{max} - r_{min}$ is bounded.  (Perhaps more general
complex paths and surfaces exist if both $r,\tau$ are complexified.)

Thus using (\ref{sphExtSur-1}), (\ref{sphExtSur-2}), 
(\ref{sphExtSur-3}), the area (\ref{RTdS01}) in $dS_{d+1}$ becomes
\be\label{sphExtSurArea}
S_{dS} = {R_{dS}^{d-1} \Omega_{d-2}\over 4G_{d+1}} 
\int_{\tau_{UV}}^{\tau_*} {d\tau\over\tau^{d-1}} (-il) 
(l^2+\tau^2)^{(d-3)/2}\ .
\ee
The integration is along the path $\tau=iT$, with $\tau_{UV}=i\epsilon$ 
and $\tau_*=il$. The leading divergence here is of the form of the 
area law (for $d>2$), given by
\be\label{Sdiv}
S_{dS}^{div}\ =\ {i \over d-2} 
{R_{dS}^{d-1} \Omega_{d-2}\over 4G_{d+1}} {l^{d-2}\over \tau_{UV}^{d-2}}\
=\ {i^{1-d}\over d-2} 
{R_{dS}^{d-1}\over 4G_{d+1}} {{\cal A}_{d-2}\over \epsilon^{d-2}}\ ,
\ee
with ${\cal A}_{d-2}\equiv l^{d-2}\Omega_{d-2}$ the interface area.
There is an overal $\pm$ sign ambiguity in the choice of the square 
root branch in (\ref{RTdS01}), (\ref{sphExtSurArea}), which we have 
fixed to be $+$ in (\ref{Sdiv}), as in the strip subregions reviewed 
earlier. This sign corresponds to choosing $\sqrt{-l^2}=-il$ 
in (\ref{sphExtSurArea}). As for the strip \cite{Narayan:2015vda}, 
the leading divergence here has the form
${\cal C}_d {{\cal A}_{d-2}\over \epsilon^{d-2}}$ with 
${\cal C}_d \sim i^{1-d} {R_{dS}^{d-1}\over G_{d+1}}$ of the form 
appearing in the $\langle TT\rangle$ correlators in 
\cite{Maldacena:2002vr}.
We also see that analytic continuation using (\ref{AdStodS}) from 
the leading area law divergence from the Ryu-Takayanagi expression 
in $AdS/CFT$ gives\ 
${R^{d-1}\over 4G_{d+1}} {l^{d-2} \Omega_{d-2}\over (d-2) \epsilon^{d-2}}\ 
\longrightarrow\ {i^{1-d}\over d-2} 
{R_{dS}^{d-1} \over 4G_{d+1}} {l^{d-2} \Omega_{d-2}\over \epsilon^{d-2}}$\ 
which is the sign above.

There are subleading terms: \eg\ for $dS_4$, the area 
(\ref{sphExtSurArea}) gives\ 
\be
S_{dS}= {R_{dS}^2\Omega_1\over 4G_4} (-il) 
\int_{\tau_{UV}}^{\tau_*} {d\tau\over\tau^2}\ 
=\ -{\pi R_{dS}^2\over 2G_4} \Big({l\over\epsilon} - 1\Big)\ .
\ee
The finite constant cutoff-independent piece ${\pi R_{dS}^2\over 2G_4}$ 
is a universal term.
For $d$ even, one of the subleading terms is the universal logarithmic 
term. Expanding (\ref{sphExtSurArea}), this logarithmic 
term can be seen to be
\be\label{logCoeffExtSurdS}
%{}^{{}^{{d-3\over 2}}}C_{{d-2\over 2}} 
%\binom{{d-3\over 2}}{{d-2\over 2}}
-i {{d-3\over 2}\choose {d-2\over 2}}\  {\Omega_{d-2}\over 4}\ 
{R_{dS}^{d-1}\over G_{d+1}}\ \log {l\over\epsilon}\ ,
\ee
where %${}^nC_k$ 
${\nu\choose k}$ is the (generalized) binomial coefficient of the 
$x^k$-term in the expansion of $(1+x)^\nu$\ 
%(note that these coefficients are not integers in general), 
and $\Omega_d = {2\pi^{(d+1)/2}\over \Gamma((d+1)/2)}$ is the $d$-dim 
sphere volume.\ The argument in the logarithmic term is obtained as 
${\tau_*\over\tau_{UV}}={il\over i\epsilon}$~.\ 
Explicitly, the coefficients (\ref{logCoeffExtSurdS}) for 
$dS_3,\ dS_5,\ dS_7$ are
\be\label{logcoeffsdS357}
-i {R_{dS}\over 2G_3} \quad [dS_3]\ ,\qquad 
-i{\pi R_{dS}^3\over 2G_5} \quad [dS_5]\ ,\qquad
-i{\pi^2 R_{dS}^5 \over 4G_7} \quad [dS_7]\ .
\ee
These coefficients resemble those arising in the $\langle TT\rangle$ 
correlators in \cite{Maldacena:2002vr}, except that the numerical 
factors are unambiguously fixed here.\ 
For $dS_3$, the area contains only the logarithmic term and the 
coefficient can be calculated directly\footnote{Explicitly the surface 
is parametrized by the two half-surfaces $x_L=-\sqrt{(l^2/4)-T^2} ,\ 
x_R=\sqrt{(l^2/4)-T^2}$ satisfying the boundary conditions $x_L\ra -l/2,\ 
x_R\ra l/2$ as $\tau=iT\ra 0$ and $x_L, x_R\ra 0$ as $\tau\ra\tau_*=il/2$. 
It is easy to check that these join smoothly at $\tau_*$: the 
resulting surface can be recognized as a continuation of the $AdS_3$ 
case. The area is $S_{dS}=2{R_{dS}\over 4G_3} \int_{i\epsilon}^{il/2} 
{d\tau\over\tau} (-il) {1\over\sqrt{l^2+\tau^2}}$ , with log-coefficient 
as above.} from (\ref{RTdS01}), (\ref{sphExtSurArea}): 
writing this area as ${c\over 3}\log {l\over\epsilon}$ gives the 
central charge $c=-i{3R_{dS}\over 2G_3}$ which can be seen to be 
the analytic continuation of the known $AdS_3$ central charge 
${3R_{AdS}\over 2G_3}$~.

Note that $-i{R_{dS}^{d-1}\over G_{d+1}}$ under the analytic
continuation (\ref{AdStodS}) becomes 
$(-1)^{{d\over 2}-1}{R_{AdS}^{d-1}\over G_{d+1}}$ which we recall arises 
in the universal coefficient of the logarithmic term in the $AdS$ case 
for spherical surfaces \cite{Ryu:2006ef}\ (and the numerical factors 
also corroborate). This coefficient is proportional to the $a$ central
charge appearing in the trace anomaly of the CFT on a sphere (for even
$d$): note that in Einstein gravity, the central charges $a, c$ are
the same, with $a\sim {R_{AdS}^{d-1}\over G_{d+1}}$ 
\cite{Henningson:1998gx,Henningson:1998ey,Balasubramanian:1999re,
de Haro:2000xn}.

This suggests that in $dS/CFT$, the coefficients of the logarithmic
term for these complex extremal surfaces are the analogs of these
$a$-type central charges in the Einstein gravity approximation.  These
coefficients match with those in the logarithmically divergent terms
in the CFT free energy evaluated using $dS/CFT$ as we discuss now.

\section{$\Psi \sim e^{iS}$,\ CFT on sphere and conformal anomaly}

For what follows, it is useful to recall the $dS/CFT$ correspondence
for de Sitter space. A version of $dS/CFT$
\cite{Strominger:2001pn,Witten:2001kn,Maldacena:2002vr} states that
quantum gravity in de Sitter space is dual to a Euclidean CFT living
on the boundary $\cI^+$.  More specifically, the CFT
partition function with specified sources $\phi_{i0}(\vx)$ coupled to
operators $\cO_i$ is identified with the bulk wavefunction of the 
universe as a functional of the boundary values of the fields dual to
$\cO_i$ given by $\phi_{i0}(\vx)$.  In the classical regime this
becomes\ $Z_{CFT} = \Psi[\phi_{i0}(\vx)] \sim e^{iS_{cl}[\phi_{i0}]}$\
where we need to impose regularity conditions on the past cosmological
horizon $\tau\ra -\infty$:\ \eg\ scalar modes satisfy
$\phi_k(\tau)\sim e^{ik\tau}$, which are Hartle-Hawking (or
Bunch-Davies) initial conditions. Operationally, certain $dS/CFT$
observables can be obtained by analytic continuation (\ref{AdStodS})
from $AdS$ (see \eg\ \cite{Maldacena:2002vr}, as well as
\cite{Harlow:2011ke}).

For even dimensions $d$, the free energy of the CFT$_d$ on a sphere is
expected to contain a logarithmic divergence whose coefficient is
related to the integrated conformal anomaly of the CFT. Since the
(nonunitary) CFT here is that dual to de Sitter space, this can be
calculated holographically using the $dS/CFT$ dictionary $Z_{CFT}=\Psi$\
\cite{Maldacena:2002vr} with an auxiliary de Sitter space in global
coordinates whose constant time slices are spheres. In the classical
regime, we approximate the Hartle-Hawking wavefunction of the universe
$\Psi$ in terms of the bulk action $S$ of this auxiliary de Sitter
space: this gives $-F=\log Z_{CFT} = \log \Psi\sim iS$.  We can then
calculate the coefficient of the logarithmic term in the classical
approximation.

de Sitter space $dS_{d+1}$ in global coordinates, with scale $R_{dS}$, is
\be\label{dSmetGlobal}
ds^2 = -dt^2 + R_{dS}^2 \Big(\cosh{t\over R_{dS}}\Big)^2 d\Omega_d^2\ .
\ee
The spatial slices are $d$-spheres, with minimum radius $R_{dS}$ at $t=0$. 
This is a solution to Einstein gravity $R_{MN}={d\over R_{dS}^2} g_{MN}$ 
with cosmological constant $\Lambda = {d(d-1)\over 2R_{dS}^2}$.\ 
The on-shell bulk action is
\be\label{dSglobAction}
S = {1\over 16\pi G_{d+1}} \int d^{d+1}x \sqrt{-g} (R-2\Lambda)
= {1\over 16\pi G_{d+1}} \int dt d^d\Omega_d\ R_{dS}^d 
\Big(\cosh{t\over R_{dS}}\Big)^d\ {2d\over R_{dS}^2}\ ,
\ee
where $R-2\Lambda = {d(d+1)-d(d-1)\over R_{dS}^2}$ and $\Omega_d$ the 
$d$-dim sphere volume. We have suppressed writing the surface terms 
and counterterms for cancelling the leading divergences in this action 
since the logarithmic term we are interested in arises solely from 
the bulk action: this is motivated by similar arguments in $AdS/CFT$ 
(see \eg\ \cite{Henningson:1998gx,Henningson:1998ey,Balasubramanian:1999re,
de Haro:2000xn} and the review \cite{Aharony:1999ti}). This gives
\be
S = {2d\ \Omega_d R_{dS}^{d-1} \over 16\pi G_{d+1}} 
\int {dt\over R_{dS}}\ \Big(\cosh{t\over R_{dS}}\Big)^d
\ =\ {R_{dS}^{d-1} \over 16\pi G_{d+1}} {2d\ \Omega_d \over 2^d}
\int d \Big({t\over R_{dS}}\Big)\  e^{dt/R_{dS}} (1 + e^{-2t/R_{dS}})^d\ .
\ee
With $\tau=-2R_{dS}e^{-t/R_{dS}}$, the metric (\ref{dSmetGlobal}) at 
asymptotically late times becomes of Poincare form\ 
$ds^2\sim {R_{dS}^2\over\tau^2} (-d\tau^2+R_{dS}^2d\Omega_d^2)$.
It is useful to write the bulk action by redefining 
$y=e^{t/R_{dS}}={2R_{dS}\over -\tau}$\ and we obtain
\be\label{dSglobAction2}
S =\ {R_{dS}^{d-1}\over 16\pi G_{d+1}} {2d\ \Omega_d \over 2^d} 
\int^{y_{UV}} dy\ y^{d-1} \Big(1 + {1\over y^2}\Big)^d\ .
\ee
The upper limit of integration here is at large $t$ \ie\ the future 
cutoff $\tau_{UV}$. The lower limit will not be important in what 
follows as long as some regularity conditions are satisfied 
%: it could be taken as the $t=0$ slice where the sphere $S^d$ shrinks 
%to minimal size (and the spacetime could be taken to be glued onto a 
%Euclidean half-sphere as in 
(see \eg\ the Hartle-Hawking prescription \cite{Hartle:1983ai}).

The expansion of this action has a logarithmic term for $d$ even.
Now the wavefunction of the universe in the classical approximation 
is\ $\Psi = e^{iS}$ and the free energy is\ 
$-F \equiv \log Z = \log\Psi \equiv iS$\ \cite{Maldacena:2002vr}.
Thus the logarithmic term in the free energy can be found by expanding 
the action (\ref{dSglobAction2}), which arises as
\be\label{logCoeffdSCFT}
iS = \ldots\ +\ i{d\choose {d\over 2}} {2d\ \Omega_d \over 16\pi\ 2^d} 
{R_{dS}^{d-1}\over G_{d+1}} \ \log {R_{dS}\over\epsilon}\ +\ \ldots\ 
=\  -i{d\choose {d\over 2}} {2d\ \Omega_d \over 16\pi\ 2^d} 
{R_{dS}^{d-1}\over G_{d+1}} \ \log\epsilon\ +\ \ldots\ ,
\ee
where the cutoff is\ $y_{UV}={2R_{dS}\over \epsilon}$\ and 
${\nu\choose k}$ is the binomial coefficient. In Euclidean $AdS_{d+1}$ 
with metric $ds^2=d\rho^2+R_{AdS}^2\sinh^2({\rho\over R_{AdS}}) d\Omega_d^2$ 
(the expected gravity dual for a conventional unitary Euclidean CFT on 
a sphere), a similar calculation yields the anomaly coefficient as is 
well known \cite{Henningson:1998gx,Henningson:1998ey,Balasubramanian:1999re,
de Haro:2000xn}: the Euclidean $AdS$ action is\ 
$S^{EAdS}={1\over 16\pi G_{d+1}} \int_{\epsilon_z} dz d^dx\sqrt{g} 
(R+2|\Lambda|)$\ (and $Z\sim e^{-S^{EAdS}}$), and the relevant terms 
arise in the expansion near the boundary. Under the analytic 
continuation $z\ra -i\tau,\ R_{AdS}\ra -iR_{dS}$, we have 
$-S^{EAdS}\ra\ iS^{dS}$ and the $-i{R_{dS}^{d-1}\over G_{d+1}}$ factor 
above continues to $(-1)^{{d\over 2}-1} {R_{AdS}^{d-1}\over G_{d+1}}$ 
in $EAdS$.\ \ Near the boundary $z=\epsilon_z$, an asymptotically 
$EAdS_5$ space 
$ds^2 = {R_{AdS}^2\over z^2} (dz^2+ {\hat g}_{\mu\nu} dx^\mu dx^\nu)$\ 
gives\ \ $-S_{EAdS_5}\sim {R_{AdS}^3\over G_5} ({\#\over\epsilon_z^4} 
- {\#\over\epsilon_z^2} - \# \log\epsilon_z + \ldots)$, with the 
$\#$ being positive coefficients for $S^4$ boundary (and 
$z=2R_{AdS}e^{-\rho/R_{AdS}}$ near the boundary). Most terms 
analytically continue to give pure imaginary terms: with 
$\epsilon_z\ra -i\epsilon_\tau$, the log-term continues as 
$\log\epsilon_z \ra \log|\epsilon_\tau| + i{\pi\over 2}$, the final 
term giving a real factor in $\Psi$ (see \cite{Maldacena:2011mk}
for interesting discussions on relations of the coefficient of 
this logarithmic term to the Hartle-Hawking factor $|\Psi|^2$). 
From the point of view of the $dS$ calculation (\ref{dSglobAction2}), 
$\tau$ being real makes the action real and so $iS$ is pure 
imaginary: a real part in $\Psi$ is obtained by deforming the contour 
slightly in the far past, \eg\ 
$y_{IR}\sim {2R_{dS}\over |\tau_{IR}|-i{\tilde\epsilon}}\sim 
i{\tilde\epsilon}$ as $\tau_{IR}\ra -\infty$. Then the logarithmic 
term gives a real term as
$iS\sim \ldots - {i\pi R_{dS}^3\over 2G_5} \log y_{IR} \sim 
- {i\pi R_{dS}^3\over 2G_5} (\log i) = {\pi^2 R_{dS}^3\over 4G_5}$ 
and correspondingly the Hartle-Hawking factor 
$|\Psi|^2\sim exp[{\pi^2 R_{dS}^3\over 2G_5}]$\ (this agrees with 
\cite{Maldacena:2011mk} using $M_{Pl}^3={1\over 8\pi G_5}$).
We also note previous work \cite{Nojiri:2001mf} on the conformal 
anomaly in $dS/CFT$ (which is however not based on the wavefunction 
of the universe).

The coefficient of this logarithmic term in (\ref{logCoeffdSCFT}) in
the free energy via $\Psi$ can be seen to be the same as that in the
logarithmic term (\ref{logCoeffExtSurdS}) in the complex spherical
extremal surfaces. They appear to be the analogs of the $a$-type
central charges in $dS/CFT$.

Further light is shed on this calculation in light of
\cite{Casini:2011kv}. A conformal mapping was used there to transform
the entanglement entropy of a spherical subsystem in $AdS/CFT$ to the
thermal entropy of the CFT in the static patch of an auxiliary de
Sitter space. For $AdS$, this allows a precise comparison with the
coefficient of the logarithmic term appearing in the extremal surface
area. These coefficients are related to the conformal anomaly and the
central charge $a$ (which is also $c$) \cite{Ryu:2006ef} in the
Einstein gravity approximation (see also
\cite{Myers:2010xs,Myers:2010tj} for higher derivative theories).

It would appear that a similar argument is at play here modulo some 
caveats (below). The CFT in this case, intrinsically Euclidean, lives 
on the flat Euclidean space on the future boundary ${\cal I}^+$ of 
de Sitter space and is nonunitary. 
We use a conformal mapping to tranform this flat $d$-dim Euclidean 
space $ds_E^2 = dt_E^2 + dr^2 + r^2 d\Omega_{d-2}^2$ to a sphere: this 
is given by the coordinate transformation
\bea
&& t_E = l {\cos\theta \sin {\rho\over l}\over 
1 + \cos\theta \cos {\rho\over l}}\ ,\qquad \qquad  
r = l {\sin\theta \over 1 + \cos\theta \cos {\rho\over l}}\ : \nonumber\\
&& ds_E^2 = \Omega^2 [\cos^2\theta d\rho^2 + l^2 (d\theta^2 + 
\sin^2\theta d\Omega_{d-2}^2)] \ ,\qquad 
\Omega={1\over 1 + \cos\theta \cos {\rho\over l}}\ .
\eea
Removing the conformal factor $\Omega^2$, this space becomes\ %\be
$d{\tilde s}^2 = \cos^2\theta d\rho^2 + l^2 (d\theta^2 + 
\sin^2\theta d\Omega_{d-2}^2)$.\  %\ee
Demanding that this space be smooth, we must avoid a conical 
singularity at $\theta={\pi\over 2}$: then the coordinate $\rho$ must 
be taken to be periodic with period $\Delta \rho = 2\pi l$.
This space can then be seen to be a $d$-sphere\
$d{\tilde s}^2 = l^2 (d\theta_1^2+\sin^2\theta_1 d\theta_2^2
+\sin^2\theta_1\sin^2\theta_2 d\Omega_{d-2}^2)$,\
using a coordinate transformation $\sin\theta=\sin\theta_1 \sin\theta_2,\ 
\tan{\rho\over l} = \cos\theta_2\tan\theta_1$.

The free energy $F$ of the Euclidean CFT on this sphere is expected to 
exhibit a logarithmically divergent term (in even dimensions) whose 
coefficient is related to the conformal anomaly. In general, we have 
an expansion 
$-F_{CFT} = \log Z_{CFT} =$\ (non-universal terms)\ $+\ a \log \epsilon\ +$\ 
(finite),\ with $\epsilon$ the ultraviolet cutoff. The CFT 
energy-momentum tensor \cite{Maldacena:2002vr} is defined as 
$T_{ij}={2\over\sqrt{h}} {\delta Z_{CFT}\over\delta h^{ij}} = 
{2\over\sqrt{h}} {\delta \Psi\over\delta h^{ij}}$ which becomes 
$T_{ij}\sim {2\over\sqrt{h}} {\delta (-F_{CFT})\over\delta h^{ij}} 
\sim i {2\over\sqrt{h}} {\delta S\over\delta h^{ij}}$ in the classical 
approximation for $\Psi$.\ Under an infinitesimal conformal 
transformation  $h_{ij}\ra (1+2\delta\lambda) h_{\mu\nu}$, \ie\ 
$\delta h^{ij} = -(2\delta\lambda) h^{ij}$, we have\ ${\delta 
F_{CFT}\over\delta\lambda} = \int d^dx\sqrt{h} \langle T{^k}_k\rangle 
+ (div)$, which is the integrated trace anomaly. Due to conformal 
invariance, this must be equivalent to simply shifting the ultraviolet 
cutoff\ $\epsilon \ra (1-\delta\lambda)\epsilon$. This gives the 
coefficient $a = \int \langle T{^k}_k\rangle$. This argument does not 
appear to require unitarity of the conformal field theory.
We have calculated this free energy holographically assuming the 
nonunitary CFT has a de Sitter gravity dual and using the $dS/CFT$ 
dictionary $Z_{CFT} = \Psi$\ \cite{Maldacena:2002vr} with an 
auxiliary de Sitter space in global coordinates (where constant 
time slices are spheres). As we have seen, we find agreement with 
the coefficients of the logarithmic terms in the complex extremal 
surfaces earlier. The fact that these coefficients are pure 
imaginary is expected from the $i$ in the relation $-F\sim iS_{bulk}$.

Finally, we expect that from the point of view of a CFT replica
calculation, the entanglement entropy is $S_{CFT}^{EE}=-\lim_{n\ra 1}
\del_n tr\rho_A^n$ where $tr\rho_A^n={Z_n\over (Z_1)^n}$ with $Z_n$
the partition function on the $n$-sheeted replica space. A scale
change is expected to be of the form $l{\del\over\del l} S_{CFT}^{EE}
\sim \int \langle T_{\mu}{^\mu}\rangle$ which is then related to the
free energy $F_{CFT}$ so that the logarithmic term coefficient in
$S_{CFT}^{EE}$ would be of the form $a \log {l\over\epsilon}$~. This
argument does not depend on unitarity. Thus if $S_{CFT}^{EE}$ is
evaluated from the bulk side as the area $S_{dS}$ of appropriate
extremal surfaces, we expect the log-coefficients to match: this is
vindicated for the complex surfaces we have been discussing.

\section{Discussion}

We have studied complex extremal surfaces for spherical subregions on
a constant boundary Euclidean time slice of the future boundary of de
Sitter space, building on \cite{Narayan:2015vda}: as in that case,
this ends up being quite different from the $AdS$ case due to sign
differences which makes the bulk quite different in structure. For
even boundary dimensions, there is a logarithmically divergent term in
the area of these surfaces whose coefficient is a universal
term. Comparing this with a corresponding coefficient (related to the
integrated conformal anomaly) in a logarithmically divergent term in
the free energy of the dual Euclidean CFT on a sphere using the
$dS/CFT$ dictionary for a dual de Sitter space in global coordinates
in a classical approximation for the wavefunction $\Psi\sim e^{iS}$,
we find agreement including numerical factors.  This coefficient is of
the form\ $-i\nu_d {R_{dS}^{d-1}\over G_{d+1}}$ where $\nu_d$ is a
real positive numerical factor. Our analysis here and in
\cite{Narayan:2015vda} has effectively enlarged the original question
of finding solutions to the extremization problem in de Sitter space
(Poincare slicing) with our boundary conditions, since the
$\tau$-parametrization being complex lies outside the original de
Sitter $\tau$-range: the eventual answers pass the checks of agreement
of various central charges based on $Z_{CFT}=\Psi$. Perhaps this
agreement is not surprising since both sides in this Einstein gravity
approximation effectively amount to analytic continuation from the
$AdS$ case (where there is agreement), but it shows consistency
between the two sides in the present case.

From the point of view of the dual Euclidean CFT, we expect that the
central charge coefficients in the extremal surfaces area (interpreted
as entanglement entropy) must match those in the dual Euclidean CFT
obtained in \cite{Maldacena:2002vr} (which are negative or imaginary):
this includes the leading area law divergence as well as subleading
universal coefficients. We would also intuitively expect finite
cutoff-independent parts which are size-dependent measures of
entanglement entropy in the CFT.  These expectations point to the
complex extremal surfaces we have been considering which exhibit these
features. The resulting analysis for these codim-2 complex extremal
surfaces in de Sitter space in the end boils down to analytic
continuation from the Ryu-Takayanagi formulation in $AdS$ (and thus
resembles known $AdS/CFT$ results with $i$s or minus signs in
appropriate places): however this was not obvious to begin
with. Perhaps the other surfaces we have discussed are also of
interest, in other contexts.

The investigations here and those in \cite{Narayan:2015vda} thus
support the idea that the areas of these complex extremal surfaces
encode entanglement entropy of the dual Euclidean CFT in $dS/CFT$,
using the formulation $Z_{CFT}=\Psi$ with $\Psi$ the wavefunction of
the universe \cite{Maldacena:2002vr}. It also suggests that in
$dS/CFT$, the coefficients of the logarithmic terms for these complex
extremal surfaces are perhaps the analogs of the $a$-type central
charges.  Relatedly it may be interesting to study analogs of
\cite{Lewkowycz:2013nqa} in the de Sitter case. It is clear however
that all our calculations are in the bulk and so cannot clearly
pinpoint the interpretation of entanglement entropy. (As an aside
however, in the 2d CFT dual to $dS_3$ with central charge $\sim
-i{R_{dS}\over G_3}$, a replica calculation of entanglement entropy
\cite{Calabrese:2004eu,Ryu:2006ef} appears to give $\sim
-i{R_{dS}\over G_3} \log {l\over\epsilon}$ under certain assumptions,
most importantly the existence of twist sector ground states.)  In
general the notion of entanglement entropy requires certain basic
assumptions on the CFT Hilbert space, most importantly the existence
of a ground state. The dual CFT for the pure de Sitter theory here
appears to have pathologies in general such as complex conformal
dimensions (the higher spin $dS/CFT$ of \cite{Anninos:2011ui} appears
better-behaved in this regard, but geometric extremal surfaces may not
be of relevance in this case).  Also the CFT is intrinsically
Euclidean, with no notion of time evolution (while the dual bulk time
direction emerges). Thus the use of a conformal transformation along
the lines of \cite{Casini:2011kv} to map the reduced density matrix to
\eg\ a thermal one appears more delicate, in such nonunitary CFTs.  It
would appear that this CFT entanglement entropy, assuming it exists,
encodes CFT correlations and is thus likely, if only indirectly, to
also encode bulk de Sitter expectation values which have intricate
connections to the dual CFT correlation functions
\cite{Maldacena:2002vr}. These issues would be interesting to explore
further.

\vspace{6mm}
%\newpage

{\footnotesize \noindent {\bf Acknowledgements:}\ \
It is a pleasure to thank Shamik Banerjee for several useful 
discussions and initial collaboration. This work is
partially supported by a grant to CMI from the Infosys Foundation.
}

%\vspace{1mm}

\end{document}